\documentclass[preprint,,tightenlines,aps,prd,showpacs,nofootinbib]{revtex4}
\usepackage{graphicx}
\usepackage{amsmath}
\usepackage{bm}

\newcommand{\fsl}[1]{#1\hspace{-1.0ex}\slash}

\begin{document}

\title{\mbox{}\\[10pt]
Exclusive Production of the $X(3872)$ in 
$B$ Meson Decay }

\author{Eric Braaten and Masaoki Kusunoki}
\affiliation{
Physics Department, Ohio State University, 
Columbus, Ohio 43210, USA}

\date{\today}
\begin{abstract}
If the recently-discovered charmonium-like state $X(3872)$ is a 
loosely-bound S-wave molecule of the charm mesons 
$D^0 \bar D^{*0}$ or $D^{*0} \bar D^0$, 
it can be produced through the weak decay 
of the $B$ meson into $D^0 \bar D^{*0} K$ or $D^{*0} \bar D^0 K$
followed by the coalescence of the charm mesons at a long-distance 
scale set by the scattering length of the charm mesons. 
The long-distance factors in the amplitude for the decay
$B \to XK$ are determined by the binding energy of $X$, 
while the short-distance factors are essentially 
determined by the amplitudes for $B \to D^0 \bar D^{*0} K$ 
and $B \to D^{*0} \bar D^0 K$ 
near the thresholds for the charm mesons.
We obtain a crude determination of the short-distance amplitudes by 
analyzing data from the Babar collaboration 
on the branching fractions for $B\to \bar D^{(*)} D^{(*)} K$
using a factorization assumption, heavy quark symmetry, and 
isospin symmetry.  The resulting order-of-magnitude estimate 
of the branching fraction for $B^+ \to X K^+$ is compatible with 
observations provided that $J/\psi \, \pi^+ \pi^-$ 
is a major decay mode of the $X$.  
The branching fraction for $B^0 \to X K^0$ 
is predicted to be suppressed by more than an order of magnitude 
compared to that for $B^+ \to X K^+$.
\end{abstract}

\pacs{12.38.-t, 12.38.Bx, 13.20.Gd, 14.40.Gx}


\maketitle


\section{Introduction}
\label{sec:intro}

The $X(3872)$ is a narrow charmonium-like resonance near 3872 MeV 
discovered by the Belle collaboration in electron-positron collisions 
through the $B$-meson 
decay $B^\pm \to X K^\pm$ followed by the decay 
$X \to J/\psi \, \pi^+\pi^-$ \cite{Choi:2003ue}. 
This state has been confirmed by the CDF \cite{Acosta:2003zx} and
D0 \cite{Abazov:2004kp} collaborations through 
its inclusive production in proton-antiproton collisions.
The discovery mode $B^\pm \to X K^\pm$ has also been confirmed 
by the Babar collaboration \cite{Aubert:2004ns}. 
The combined measurement of the mass of the $X$ is $3871.9\pm 0.5$ MeV 
\cite{Olsen:2004fp}.
The width of the $X$ is less than 2.3 MeV at 90\% C.L.~\cite{Choi:2003ue}, 
which is much narrower than other charmonium states 
above the $D \bar D$ threshold. 
The product of the branching fractions associated with the discovery 
channel has been measured by the Belle and Babar collaborations: 
\cite{Choi:2003ue,Aubert:2004ns,Olsen:2004fp} 
\begin{eqnarray}
{\rm Br}[B^+\to XK^+] \, {\rm Br}[X \to J/\psi \, \pi^+ \pi^-] 
= (1.3 \pm 0.3) \times 10^{-5} .
\label{Brexp}
\end{eqnarray}
%
The Belle collaboration recently observed the $X(3872)$ in a
second decay mode: $X \to J/\psi \, \pi^+\pi^-\pi^0$ \cite{Abe:2004sd}. 
The invariant mass distribution of the three pions 
is dominated by a virtual $\omega$ resonance.
The branching ratio relative to the discovery decay channel is
\cite{Abe:2004sd}
\begin{eqnarray}
{{\rm Br}[X \to J/\psi  \, \omega] 
\over {\rm Br}[X \to J/\psi  \, \pi^+ \pi^-]} =
0.8 \pm 0.3_{\rm stat} \pm 0.1_{\rm syst}.
\label{Br:omegapsi}
\end{eqnarray}
%
Upper limits have been placed on the branching fractions for
other decay modes of the $X$, including
$D^0 \bar D^0$, $D^+ D^-$, $D^0 \bar D^0 \pi^0$ \cite{Abe:2003zv},
$\chi_{c1} \gamma$, $\chi_{c2} \gamma$, $J/\psi \, \gamma$,
$J/\psi \, \pi^0 \pi^0$ \cite{Abe:2004sd},
and $J/\psi \, \eta$ \cite{Aubert:2004fc}.
Upper limits have also been placed on the partial widths for the
decay of $X(3872)$ into $e^+e^-$ \cite{Yuan:2003yz,Metreveli:2004px}
and into $\gamma \gamma$ \cite{Metreveli:2004px}.

The most plausible interpretations of the $X(3872)$ are a
charmonium state with constituents $c \bar c$
\cite{Barnes:2003vb,Eichten:2004uh,Quigg:2004vf}
or a hadronic molecule with constituents $D D^*$
\cite{Tornqvist:2003na,Voloshin:2003nt,Wong:2003xk,%
Braaten:2003he,Swanson:2003tb,Tornqvist:2004qy}.
Other proposed interpretations include an S-wave threshold 
enhancement in $D^0 \bar D^{*0}$ scattering \cite{Bugg:2004sh},
a ``hybrid charmonium'' state with constituents $c \bar c g$ 
\cite{Li:2004st},
a vector glueball with a small admixture of charmonium states
\cite{Seth:2004zb},
and a diquark-antidiquark bound state with constituents 
$c u \bar c \bar u$ \cite{Maiani:2004vq}.
Measurements of the decays of the $X$ can be used to determine 
its quantum numbers and narrow down these options
\cite{Close:2003sg,Pakvasa:2003ea,Rosner:2004ac,Ko:2004cz}.
The charmonium options include members of the multiplet of 
the first radial excitation of P-wave charmonium 
and the multiplet of ground-state D-wave charmonium. 
The $h_{c}(2P)$, $\chi_{c0}(2P)$, and $\psi_1(1D)$, 
whose respective $J^{PC}$ quantum numbers are $1^{+-}$, $0^{++}$,
and $1^{--}$, have already been ruled out as candidates for $X$.
Evidence disfavoring each of the remaining states
has been steadily accumulating \cite{Quigg:2004vf,Olsen:2004fp}.

The option of a $D D^*$ molecule is motivated by the proximity 
of the $X$ to the threshold for $D^0 \bar D^{*0}$:
$m_{D^0}+ m_{D^{*0}} = 3871.3 \pm 1.0$ MeV.%
\footnote{The uncertainty in $m_{D^{*0}}-m_{D^0}$ 
is only 0.07 MeV, so the uncertainty in the threshold energy 
comes almost entirely from the 0.5 MeV uncertainty in $m_{D^0}$.}
If $X$ is a $D^0 \bar D^{*0}$/$D^{*0} \bar D^0$ molecule,
its binding energy $E_b = m_{D^0}+ m_{D^{*0}} -m_X$ is
$E_b = -0.6 \pm 1.1$ MeV.  
The central value corresponds to a resonance,
but the error bar allows a bound state with $E_b>0$.
The possibility that charm mesons might form molecular states 
was first considered some time ago 
\cite{Bander:1975fb,Voloshin:ap,DeRujula:1976qd,Nussinov:1976fg}.
Tornqvist pointed out that $D D^*$ molecules were likely 
to be very loosely bound \cite{Tornqvist:1993ng}.
If the binding of $D^0 \bar D^{*0}$ or $D^{*0} \bar D^0$ 
is due to pion exchange, the most favorable channels are
S-wave with quantum numbers $J^{PC}=1^{++}$ and P-wave with 
quantum numbers $0^{-+}$
\cite{Tornqvist:1993ng,Tornqvist:2003na,Tornqvist:2004qy}. 
In a model by Swanson that includes pion exchange 
at long distances and quark exchange at short distances, 
the only bound state is in the $1^{++}$ channel \cite{Swanson:2003tb}.
Another mechanism for generating a $D D^*$ molecule is the 
accidental fine-tuning of the mass of the 
$h_{c}(2P)$ or $\chi_{c1}(2P)$ to the $D D^*$ threshold
which creates a molecule with quantum numbers
$1^{+-}$ or $1^{++}$, respectively \cite{Braaten:2003he}.

In the decay $X \to J/\psi \, \pi^+\pi^- \pi^0$, 
the invariant mass distribution of the three pions is dominated
by a virtual $\omega$ resonance. 
In the decay $X \to J/\psi \, \pi^+\pi^-$, 
the invariant mass distribution of the two pions seems to peak 
near the upper endpoint \cite{Abe:2004sd}, which suggests that 
they come from the decay of a virtual $\rho^0$ resonance. 
If the observed decay modes have been correctly interpreted as 
$X \to J/\psi \, \rho^*$ and $X \to J/\psi \, \omega^*$,
the roughly equal branching fractions in Eq.~(\ref{Br:omegapsi}) 
implies large isospin violations.
This immediately rules out all charmonium options 
with the exceptions of $h_{c}(2P)$ and $\chi_{c1}(2P)$.
These charmonium states are exceptional, because if either of 
their masses was somehow tuned
sufficiently close to the $D^0 \bar D^{*0}$ threshold, the resonant 
interactions with S-wave $D D^*$ scattering states would 
transform the charmonium state into a state that is 
predominantly a $D D^*$ molecule.
If the $X(3872)$ is a loosely-bound $D D^*$ molecule, 
large isospin violations are expected.  They arise simply from
the fact that the mass of the $X$ is much closer to the 
$D^0 \bar D^{*0}$ threshold than to the 
$D^+ D^{*-}$ threshold at $3879.4 \pm 1.0$ MeV.

The interpretation of $X(3872)$ as a loosely-bound $D D^*$ molecule 
has important implications for the production of the $X(3872)$
\cite{Braaten:2004rn,Braaten:2004fk,Braaten:2004jg,Voloshin:2004mh}.
If the $X$ is a loosely-bound S-wave 
$D^0 \bar D^{*0}$/$D^{*0} \bar D^0$ 
molecule, it can be produced in any high-energy collision that 
can create $D^0$ and $\bar D^{*0}$ or $D^{*0}$ and $\bar D^0$. 
If these charm mesons are produced with sufficiently small 
relative momentum, they can subsequently coalesce into the $X$.
In Ref.~\cite{Braaten:2004rn}, this coalescence mechanism 
was applied to the exclusive decay process
$\Upsilon(4S)\to X h^+ h^-$, 
where $h^+$ and $h^-$ are light hadrons. 
This decay can proceed through the decay of $\Upsilon(4S)$ into
a virtual $B^+$ and a virtual $B^-$, followed by the decays 
$B^+ \to \bar D^0 h^+$ and $B^- \to D^{*0} h^-$ 
or by the decays  
$B^+ \to \bar D^{*0} h^+$ and $B^- \to D^0 h^-$,
and then finally by the coalescence of the charm mesons into the $X$. 
Remarkably, the rate for this process can be calculated in terms of 
hadron masses and the width of the $B$ meson only.
Unfortunately, it is many orders of magnitude 
too small to ever be observed in an experiment. 
The coalesence mechanism was applied to the discovery mode 
$B^+ \to X K^+$ in Ref.~\cite{Braaten:2004fk}. 
An order-of-magnitude estimate of the branching fraction
was found to be consistent with current experimental 
observations if the charge conjugation quantum number of the $X$ is 
$C=+$ and if $J/\psi \, \pi^+\pi^-$ is one of its major decay modes.
It was pointed out that the molecular interpretation of the $X$ 
can be confirmed by the observation of 
a peak in the $D^0 \bar D^{*0}$ 
invariant mass distribution just above the $D^0 \bar D^{*0}$ threshold 
in the decay $B^+ \to D^0 \bar D^{*0} K^+$. 

In this paper, we present a more thorough treatment of the
coalesence mechanism for the production of $X$ in the exclusive
decays $B \to X K$.  We extend our previous work to the decay mode 
$B^0 \to X K^0$, which has not yet been observed. 
Our analysis indicates that the branching fraction for 
$B^0\to X K^0$ should be suppressed relative to that for
$B^+\to X K^+$ by more than an order of magnitude. 
A measurement or upper bound on the branching fraction 
for $B^0\to X K^0$ that is much smaller than that for $B^+\to X K^+$ 
would support the identification 
of the $X(3872)$ as a $D D^*$ molecule.

\section{Universality and the $\bm{X(3872)}$}
\label{sec:universality}

The measurement of the mass of the $X$ indicates that 
its binding energy is likely to be within 1 MeV of
the $D^0 \bar D^{*0}$ threshold.
This binding energy is small compared to the natural energy scale
associated with pion exchange: $m_\pi^2/(2\mu_{DD^*}) \approx 10$ MeV, 
where $\mu_{DD^*}$ is the reduced mass of the $D^0$ and $D^{*0}$. 
The unnaturally small binding energy $E_b$ implies that 
the S-wave $D^0 \bar D^{*0}$ scattering
length is large compared to the natural scale $1/m_\pi$.
We assume that the $D^0 \bar D^{*0}$/$D^{*0} \bar D^0$ system has a large 
scattering length $a$ in the $C=+$ channel
and that the scattering length in the $C=-$ channel 
is negligible in comparison. 
The scattering lengths for elastic $D^0 \bar D^{*0}$ scattering 
and for elastic $D^{*0} \bar D^0$ scattering are then both $a/2$.%
\footnote{In previous papers 
\cite{Braaten:2003he,Braaten:2004jg,Braaten:2004fk,Braaten:2004rn}, 
we also denoted the scattering length in the $C=+$ channel by $a$, 
but we referred to it incorrectly as the 
$D^0 \bar D^{*0}$ scattering length.}
The large scattering length $a$ could be tuned to $+\infty$ 
by adjusting a short-distance parameter in QCD.
One possible choice for this parameter is the mass of the up quark,
since the precise value of the $D^0 \bar D^{*0}$ threshold
energy depends on this mass.  
Nonrelativistic few-body systems with short-range interactions 
and a large scattering length have universal properties 
that depend on the scattering length but are otherwise insensitive 
to details at distances small compared to $a$ \cite{Hammer:2004rn}.
Thus, the $D^0 \bar D^{*0}$/$D^{*0} \bar D^0$ system
will have universal properties that are insensitive to 
any of the shorter distance scales of QCD.

If the scattering length $a$ is large and positive,
the simplest prediction of universality is that there is a 
loosely-bound 2-body bound state.
In the case of the $D^0 \bar D^{*0}$/$D^{*0} \bar D^0$ system,
this bound state can be identified with the $X(3872)$.
The universal formula for its binding energy is
\begin{eqnarray}
E_b = \frac{1}{2\mu_{DD^*} a^2}.
\label{Eb}
\end{eqnarray}
%
The reduced mass $\mu_{DD^*}$ is well-approximated by
\begin{eqnarray}
\mu_{DD^*} \simeq {m_{D^0} m_{D^{*0}} \over m_X}.
\label{mu}
\end{eqnarray}
%
The wavefunction of the $X$ for $D D^*$ separations 
$r \gg 1/m_\pi$ is universal. The normalized wavefunction is 
\begin{eqnarray}
\psi(r) = (2 \pi a)^{-1/2} {\exp(-r/a) \over r} .
\label{wfr}
\end{eqnarray}  
%
Its Fourier transform is
\begin{eqnarray}
\tilde \psi(q) = \frac{(8\pi/a)^{1/2}}{q^2 + 1/a^2}. 
\label{wfk}
\end{eqnarray}  
%
The momentum-space wavefunction has this universal form 
for $q \ll m_\pi$.

In the measurement of the binding energy,
$E_b = -0.6 \pm 1.1$ MeV, the central value is negative,
corresponding to a resonance in $D^0 \bar D^{*0}$ 
or $D^{*0} \bar D^0$ scattering rather than a bound state.  
The small negative binding energy requires a
large negative scattering length.
If this resonance was indeed the $X(3872)$, its dominant decay modes 
would be $D^0 \bar D^{*0}$ and $D^{*0} \bar D^0$.
After the subsequent decay of the $\bar D^{*0}$ or $D^{*0}$,
the ultimate final state is $D^0 \bar D^0 \pi^0$ 
or $D^0 \bar D^0 \gamma$.  The Belle collaboration has set 
an upper limit on the product of the branching fractions for 
$B^+ \to X K^+$ and $X \to D^0 \bar D^0 \pi^0$ \cite{Abe:2003zv}.
Dividing by the central value of the
corresponding product of branching fractions 
for the discovery mode, we obtain the limit
\begin{eqnarray}
{{\rm Br}[X \to D^0 \bar D^0 \pi^0] \over 
	{\rm Br}[X \to J/\psi \pi^+ \pi^-]} < 5.
\end{eqnarray}
%
This loose upper bound seems to be the only quantitative
evidence against the identification 
of $X(3872)$ as a $D^0 \bar D^{*0}$/$D^{*0} \bar D^0$ resonance.

Throughout most of this paper, we will assume that $X$ 
is a $D^0 \bar D^{*0}$/$D^{*0} \bar D^0$ bound state,
which requires that $a$ be large and positive.
The state of the $X$ can be written schematically as
\begin{eqnarray}
| X \rangle = 
{Z^{1/2} \over \sqrt{2}} 
\left( | D^0 \bar D^{*0} \rangle +  | D^{*0} \bar D^0 \rangle \right)
+ \sum_H Z_H^{1/2} |H\rangle, 
\label{X}
\end{eqnarray}
%
where $Z$ is the probability for the $X$ to be in the 
$D^0 \bar D^{*0}$/$D^{*0} \bar D^0$ state, $Z_H$ is the probability
for the $X$ to be in another hadronic state $H$, and $Z + \sum_H Z_H = 1$. 
The other hadronic states $H$ could include charmonium states
such as $\chi_{c1}(2P)$,
scattering states of charm mesons such as $D^+ D^{*-}$,
and scattering states of a charmonium and a light hadron
such as $J/\psi \, \rho$ or $J/\psi \, \omega$.
Universality implies that $Z_H$ scales like $1/a$ 
and $Z$ approaches 1 as $a$ increases \cite{Braaten:2003he}. 
In the limit $a\to \infty$, the state is a pure
$D^0 \bar D^{*0}$/$D^{*0} \bar D^0$ molecule.  

The amplitudes for the scattering of $D^0 \bar D^{*0}$ 
(or $\bar D^0 D^{*0}$) with sufficiently small relative momentum
are universal \cite{Braaten:2003he}.
If the $D^0$ and $\bar D^{*0}$ have momenta $\pm {\bm q}$
in the $D D^*$ rest frame with $|{\bm q}| \ll m_\pi$,
the universal expressions for their amplitudes 
to scatter into $D^0 \bar D^{*0}$ and $D^{*0} \bar D^0$ are
\begin{subequations}
\begin{eqnarray}
{\cal A}[D^0 \bar D^{*0} \to D^0 \bar D^{*0}] &=& 
{8 \pi m_X \over -1/a - i q},
\\
{\cal A}[D^0 \bar D^{*0} \to D^{*0} \bar D^0] &=& 
{8 \pi m_X \over -1/a - i q}.
\end{eqnarray}
\label{amp-scat}
\end{subequations}
%
We have implicitly assumed that the $D^*$ has the same polarization 
vector in the initial and final states.
The numerator in Eqs.~(\ref{amp-scat}) differs from the 
product of the standard nonrelativistic factor $4 \pi/\mu_{D D^*}$ 
and the relativistic normalization factor $4 m_D m_{D^*}$ 
by a factor of $(1/\sqrt{2})^2$ from the projection 
of the initial and final $D D^*$ states onto the $C=+$ channel.
The amplitudes for  $D^0 \bar D^{*0}$ or $D^{*0} \bar D^0$ 
with $|{\bm q}| \ll m_\pi$ to coalesce into $X$
are also universal \cite{Braaten:2003he}:
\begin{subequations}
\begin{eqnarray}
{\cal A}[D^0 \bar D^{*0} \to X] &=& 
(16\pi m_X^2/\mu_{DD^*} a)^{1/2},
\\
{\cal A}[D^{*0} \bar D^0 \to X] &=& 
(16\pi m_X^2/\mu_{DD^*} a)^{1/2}.
\end{eqnarray}
\label{amp-coal}
\end{subequations}
%
We have implicitly assumed that the $X$ has the same polarization vector 
as the $D^*$.
Otherwise these amplitudes have an additional factor of
$\epsilon_X^* \cdot \epsilon_{D^*}$.
We have simplified the expressions for the amplitudes given 
in Ref.~\cite{Braaten:2004rn}
by setting $Z=1$ and using the expression 
for the reduced mass in Eq.~(\ref{mu}).

Any production process for $X$ necessarily involves a wide range
of important momentum scales.
In the decay $B \to X K$,
the largest momentum scale is the mass $M_W \approx 80$ GeV
of the $W$ boson that mediates the quark decay process
$b \to c \bar c s$.  
The next largest momentum scale is $m_b - 2 m_c \approx 1.5$ GeV, 
which is the scale of the energy of the $s$ 
that recoils against the $c \bar c$ system.
Then there is the scale $\Lambda_{\rm QCD} \approx 300$ MeV
associated with the wavefunctions of light quarks in the hadrons
and the scale $m_\pi \approx 140$ MeV associated 
with the pion-exchange interaction between the charm mesons.
The smallest momentum scale is set by the 
wavefunction of the $D D^*$ molecule: 
$1/a < 30$ MeV if $E_b < 0.5$ MeV.
This hierarchy of momentum scales can be summarized by the 
inequalities
\begin{eqnarray} 
1/a \ll m_\pi < \Lambda_{\rm QCD} \ll m_b - 2 m_c \ll M_W.
\label{hierarchy}
\end{eqnarray}
%
The two largest momentum scales, whose ratio is about 50, 
provides a very large hierarchy.  It can be exploited by
replacing the effects of the virtual $W$ 
by an effective weak hamiltonian that involves local 
interactions between quark fields.
The two smallest momentum scales in Eq.~(\ref{hierarchy})
may also provide a very large hierarchy.
The ratio $m_\pi a$ is greater than 4.3 if $E_b<0.5$ MeV
and it could be much greater.

We can exploit the universality of nonrelativistic particles 
with large scattering lengths by introducing an ultraviolet cutoff 
$\Lambda$ on the relative momentum ${\bm q}$ of the 
$D^0$ or $\bar D^0$ in the $D D^*$ rest frame.
If we choose this cutoff in the range
\begin{eqnarray}
1/a \ll \Lambda \ll m_\pi,
\end{eqnarray}
%
the behavior of $D^0 \bar D^{*0}$ or $D^{*0} \bar D^0$ 
with $|{\bm q}| < \Lambda$ is governed by universality.  
We will refer to processes involving such small momenta 
as {\it long-distance}, while processes involving momenta satisfying 
$|{\bm q}| > \Lambda$ will be referred to as {\it short-distance}.
In the next two sections, we apply this separation of scales to 
the decays $B \to D^0 \bar D^{*0} K$ and $B \to X K$.
 
The ultraviolet momentum cutoff $\Lambda$ implies an ultraviolet 
energy cutoff $\Lambda^2/(2 \mu_{D D^*})$.  In the decomposition
of $X$ in Eq.~(\ref{X}), all states $H$ for which the energy
gap $|M_H - (m_{D^0} + m_{D^{*0}})|$ is greater than 
$\Lambda^2/(2 \mu_{D D^*})$ can be excluded.  The effects of such 
highly virtual states can be taken into account indirectly 
through the scattering length and through other 
effects on the $D D^*$ states.  If $a$ is large enough,  
we can choose the ultraviolet energy cutoff 
$\Lambda^2/(2 \mu_{D D^*})$ to be smaller than the 
smallest energy gap.  In this case,
only the $D^0 \bar D^{*0}$ and $D^{*0} \bar D^0$ states in
the decomposition in Eq.~(\ref{X}) remain and we can set $Z=1$.

The large $D D^*$ scattering length could arise from a fortuitous 
fine-tuning of the pion-exchange potential between $D^0$ 
and $\bar D^{*0}$ and between $D^{*0}$ and $\bar D^0$
in the $C=+$ channel.  If $a>0$, there is necessarily a bound state 
near the $D^0 \bar D^{*0}$ threshold, and it is identified 
with the $X$.  Alternatively, the large scattering length 
could arise from a {\it Feshbach resonance} \cite{Hammer:2004rn}, 
which requires the 
fine-tuning of the potential in a closed channel so that 
a bound state in that channel is close to the 
$D^0 \bar D^{*0}$ threshold.
If the closed channel is weakly coupled to $D^0 \bar D^{*0}$ and 
$D^{*0} \bar D^0$ in the absence of the fine tuning, then  
the fine tuning generates a large scattering length
when the bound state is extremely close to the $D^0 \bar D^{*0}$ 
threshold.  The natural choice for the closed channel is 
$c \bar c$, in which case the bound state in the closed channel 
could be the P-wave charmonium state $\chi_{c1}(2P)$.
Since the decay of the $X(3872)$ seems to proceed through 
$J/\psi \, \rho^*$ and $J/\psi \, \omega^*$, one might be tempted 
to identify the closed channel responsible for the Feshbach 
resonance with $J/\psi \, \rho$ or $J/\psi \, \omega$.
However the widths of the particles in the closed channel 
provides a lower bound on the width of the Feshbach resonance.  
The identification of the closed channel with $J/\psi \, \rho$ 
or $J/\psi \, \omega$ is therefore excluded by the fact that the 
widths of $\rho$ and $\omega$ are larger than the experimental 
upper bound on the width of the $X(3872)$.

\section{The Decay $\bm{B \to D^0 {\bar D}^{*0} K}$}
\label{sec:DDK}

\begin{figure}
\includegraphics[width=14cm]{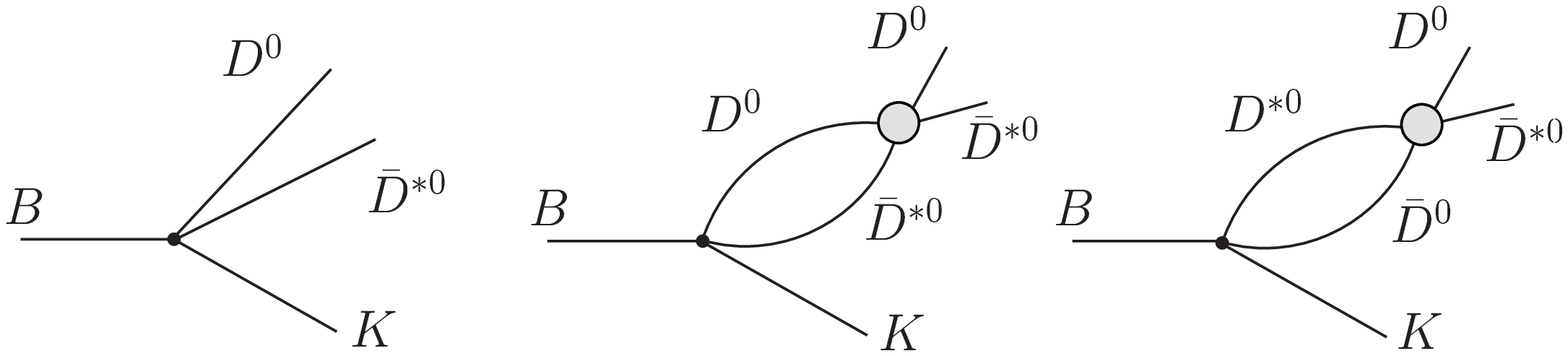}
\caption{
Feynman diagrams for the decay $B \to D^0 \bar D^{*0} K$,
with the short-distance decay amplitudes represented by dots 
and the long-distance scattering amplitudes represented by blobs.
\label{fig:B-DDK}}
\end{figure}

We proceed to apply the separation of the long-distance scale $a$
from the shorter distance scales of QCD to the decay process 
$B \to D^0 \bar D^{*0} K$. 
We denote the 4-momenta of the $B$, $D^0$, $\bar D^{*0}$, and $K$
by $P$, $q$, $q_*$, and $k$, respectively.
We take the relative 3-momentum ${\bm q}= -{\bm q}_*$ 
in the $D^0 \bar D^{*0}$ rest frame to be smaller than the 
separation scale $\Lambda$. 
The amplitude for $B \to D^0 \bar D^{*0} K$ can be decomposed into 
three terms corresponding to the three diagrams in Fig.~\ref{fig:B-DDK}.
The first diagram represents the direct production of 
$D^0 \bar D^{*0}$ by the decay $B \to D D^* K$ at short distances.
The second diagram represents the decay 
$B\to D^0 \bar D^{*0} K$  at short distances followed by the 
elastic scattering of $D^0 \bar D^{*0}$ at long distances.
The third diagram represents the decay 
$B\to D^{*0} \bar D^0 K$ at short distances followed by the 
scattering of $D^{*0} \bar D^0$ into $D^0 \bar D^{*0}$
at long distances.
The expression for the amplitude is
\begin{eqnarray}
{\cal A}[B \to D^0 \bar D^{*0} K] &=& 
{\cal A}_{\rm short}[B \to D^0 \bar D^{*0} K]
\nonumber
\\
&&  - i  \int \! \! \frac{d^4\ell}{(2\pi)^4} \, 
{\cal A}_{\rm short}[B \to  D^0 \bar D^{*0} K] 
\nonumber
\\
&& \hspace{1.5cm} \times 
D(q+\ell,m_{D^0}) \, D(q_*-\ell,m_{D^{*0}}) \, 
{\cal A}[ D^0 \bar D^{*0} \to D^0 \bar D^{*0}  ]
\nonumber
\\
&&  - i  \int \! \! \frac{d^4\ell}{(2\pi)^4} \, 
{\cal A}_{\rm short}[B \to D^{*0} \bar D^0 K] 
\nonumber
\\
&& \hspace{1.5cm} \times 
D(q+\ell,m_{D^0}) \, D(q_*-\ell,m_{D^{*0}}) \, 
{\cal A}[ D^{*0} \bar D^0 \to D^0 \bar D^{*0} ].
\label{ampDD-sum}
\end{eqnarray}
%
The propagators of the virtual $D$ and $D^*$ are
\begin{eqnarray}
D(p,m) = (p^2 - m^2 + i \epsilon)^{-1}.
\end{eqnarray}
%
In the loop integrals, there is an implicit ultraviolet cutoff 
$|{\bm \ell}| < \Lambda$ on the 3-momenta of the virtual
$D$ and $D^*$ in the $D D^*$ rest frame.

The long-distance scattering amplitudes in Eqs.~(\ref{ampDD-sum})
are given by the universal expressions in Eqs.~(\ref{amp-scat}).
In the short-distance decay amplitudes in Eq.~(\ref{ampDD-sum}),
we can neglect the relative 3-momentum ${\bm \ell}$ 
of the $D$ and $D^*$, since it is small 
compared to all the other momenta in the process.
The 4-momenta of the $D$ and $D^*$ are well-approximated
by $(m_{D^0}/m_X) Q^\mu$ and $(m_{D^{*0}}/m_X) Q^\mu$,
where $Q = P - k$.
Lorentz invariance then constrains the short-distance decay amplitudes
to have the very simple forms
\begin{subequations}
\begin{eqnarray}
 {\cal A}_{\rm short}[B \to  D^0 \bar D^{*0} K]
&=& c_1(\Lambda) \, P \cdot \epsilon^*,
\label{c1c2:a}
\\
 {\cal A}_{\rm short}[B\to  D^{*0} \bar D^0 K]
&=& c_2(\Lambda)  \, P \cdot \epsilon^*,
\label{c1c2}
\end{eqnarray} 
\label{amp-DDK}
\end{subequations}
%
where $\epsilon$ is the polarization 4-vector of the $D^*$. 
The coefficients $c_1$ and $c_2$, which depend on the separation 
scale $\Lambda$, have dimensions of inverse energy.

In the loop integrals in Eq.~(\ref{ampDD-sum}), the integral over 
the variable $\ell_0$ can be evaluated by applying the residue theorem
to the appropriate pole in one of the $D$ meson propagators:
\begin{eqnarray}
\int \frac{d^4\ell}{(2\pi)^4} D(q+\ell,m_{D^0}) \, D(q_*-\ell,m_{D^{*0}})
={i \over 2 m_X} \left( {a \over 8 \pi} \right)^{1/2}
\int \frac{d^3\ell}{(2\pi)^3} \tilde \psi(\ell),
\label{int-ell0}
\end{eqnarray} 
%
where $\tilde \psi(\ell)$ is the universal wavefunction of the $X$
given in Eq.~(\ref{wfk}).  The remaining integral 
must be evaluated using the ultraviolet cutoff 
$|{\bm \ell}| < \Lambda$:
\begin{eqnarray}
\int \frac{d^3 \ell}{(2\pi)^3} \tilde \psi(\ell)
= \left( {2 \over \pi^3 a} \right)^{1/2}
\left[ \Lambda - {\arctan(a \Lambda) \over a} \right].
\label{int-psi}
\end{eqnarray} 
%
Putting all the ingredients together 
and keeping only the leading terms for $a \Lambda \gg 1$ in each of the 
three contributions, the decay amplitude in Eq.~(\ref{ampDD-sum}) 
reduces to
\begin{eqnarray}
{\cal A}[B \to  D^0 \bar D^{*0} K] \simeq
\left( c_1(\Lambda) - 
	{2 a \Lambda [c_1(\Lambda) + c_2(\Lambda)] \over \pi (1 + i a q)} \right) 
P \cdot \epsilon^*.
\label{amp-XK:Lam}
\end{eqnarray} 
%

\begin{figure}
\includegraphics[width=12cm,angle=270]{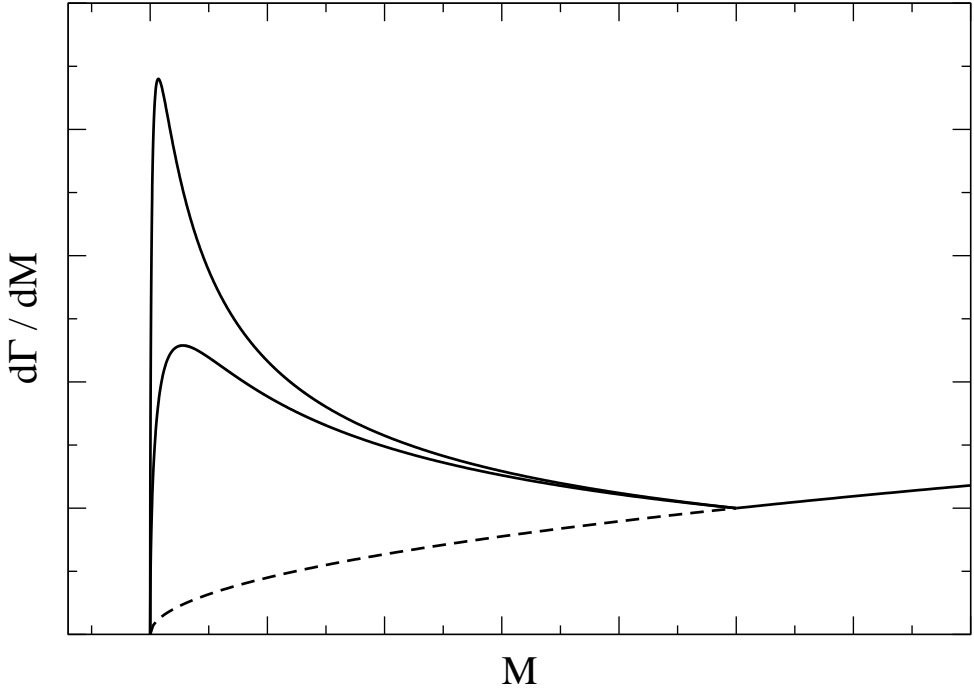}
\caption{Invariant mass distribution  
for $D^0 \bar D^{*0}$ near threshold 
(solid lines) for two values of the large scattering length 
that differ by a factor of 2.  The peak in the invariant mass 
$M = m_{D^0} + m_{D^{*0}} + q^2/(2 \mu_{DD^*})$ occurs at $q = 1/a$.
The crossover from the universal curves to the
phase space distribution (dashed line) 
has been modeled by a sudden transition at $q = \Lambda_\pi$.
\label{fig:invmass}}
\end{figure}

Our original expression for the decay amplitude in 
Eq.~(\ref{ampDD-sum}) simply corresponds to a separation of scales,
so it is necessarily independent of the arbitrary scale $\Lambda$.  
However we have approximated the long-distance scattering amplitudes 
by ignoring terms suppressed by $1/(a \Lambda)$,
in which case they reduce to the 
universal expressions in Eqs.~(\ref{amp-scat}).  We have also 
approximated the short-distance amplitudes in Eqs.~(\ref{amp-DDK})
by ignoring terms suppressed by $\ell/\Lambda$.  Thus we can expect our
expression for the amplitude in Eq.~(\ref{amp-XK:Lam}) to be independent 
of $\Lambda$ only up to terms that are suppressed by powers of 
$1/\Lambda$.  This is possible only if the coefficients $c_i(\Lambda)$
scale like $1/\Lambda$, which implies that 
the combinations $\Lambda c_i(\Lambda)$
must approach ultraviolet fixed points $\hat c_i$ as $\Lambda$ increases.  
The approach to the fixed points is of course ultimately interrupted 
by the physical scale $m_\pi$.  Neglecting terms that are suppressed 
by powers of $1/\Lambda$, the decay amplitude reduces to
\begin{eqnarray}
{\cal A}[B \to  D^0 \bar D^{*0} K] =
- {2(\hat c_1 + \hat c_2) a \over \pi (1 + i a q)}  \, 
P \cdot \epsilon^*.
\label{amp-XK}
\end{eqnarray} 
%

The resulting expression for the differential decay rate 
with respect to the $D D^*$ invariant mass $M$ is 
\begin{eqnarray}
{d \Gamma \over d M} [B \to  D^0 \bar D^{*0} K] =
|\hat c_1 + \hat c_2|^2 
{\lambda^{3/2}(m_B,m_X,m_K) \over 64 \pi^5 m_B^3 m_X^2} \, 
{a^2 q \over 1 + a^2 q^2} ,
\label{dGam}
\end{eqnarray} 
%
where $q$ is the momentum of the $D$ or $D^*$ in the $D D^*$ rest frame,
\begin{eqnarray}
q = {\lambda^{1/2}(M,m_{D^0}, m_{D^{*0}}) \over 2M},
\end{eqnarray} 
%
and $\lambda(x,y,z)$ is the triangle function:
\begin{eqnarray}
\lambda(x,y,z)=x^4+y^4+z^4-2(x^2y^2+y^2z^2+z^2x^2).
\end{eqnarray} 
%
Although we assumed $a>0$ in the derivation, our final result 
in Eq.~(\ref{dGam}) is valid for either sign of $a$.
If $a<0$, the only difference in the derivation is that the
integral in Eq.~(\ref{int-ell0}) cannot be interpreted in terms 
of a universal bound-state wavefunction, since there is no bound state. 
Near the threshold, the $D D^*$ invariant mass can be approximated by
\begin{eqnarray}
M \simeq m_{D^0} + m_{D^{*0}} + q^2/(2 \mu_{DD^*}).
\end{eqnarray} 
%
The invariant mass distribution in Eq.~(\ref{dGam}) has a peak at 
$q = 1/|a|$ with a height that scales like $|a|$ as $|a|$ increases.  
It decreases to half the maximum at $q = (2 \pm \sqrt{3})/|a|$.
The full width in $M$ at half maximum is $4 \sqrt{3}/(\mu_{D D^*} a^2)$.

The decay rate for $B \to D^{*0} \bar D^0 K$ also proceeds 
by the Feynman diagrams in Fig.~\ref{fig:B-DDK},
except that the charm mesons in the final state are
$D^{*0}$ and $\bar D^0$.
The differential decay rate for $B \to D^{*0} \bar D^0 K$ 
in the scaling region $q \ll m_\pi$ is given by exactly the same 
universal expression in Eq.~(\ref{dGam}).
If the large scattering length occurred in the channel 
with charge conjugation $C = -$, the only difference 
would be that the factor $|\hat c_1 + \hat c_2|^2$ in Eq.~(\ref{dGam})
would be replaced by $|\hat c_1 - \hat c_2|^2$.
We have not specified the charge of the $B$ meson.  
The fixed-point coefficients $\hat c_1$ and $\hat c_2$ 
in Eq.~(\ref{dGam}) have different values for the decays
$B^+ \to D^0 \bar D^{*0} K^+$ and $B^0 \to D^0 \bar D^{*0} K^0$.

The expression for the differential decay rate in Eq.~(\ref{dGam}) 
applies only in the scaling region $q \ll m_\pi$.  
At larger values of $q$ that are still small compared to the scale
$m_b - 2 m_c$, the resonant terms disappear and the decay amplitude 
reduces to the short-distance 
term $c_1(\Lambda) P \cdot \epsilon^*$ in Eqs.~(\ref{c1c2:a}). 
The corresponding invariant mass distribution $d \Gamma/d M$
for $q$ just above the scaling region
follows the phase space distribution,
which is proportional to $q$ in the limit $q \to 0$.  
The crossover from the resonant distribution proportional to 
$a^2q/(1+a^2 q^2)$ to the phase space distribution proportional to $q$
occurs at a momentum scale that we will denote by $\Lambda_\pi$.
We expect $\Lambda_\pi$ to be comparable to $m_\pi$.
Just above the crossover region, 
the differential decay rate can be approximated by
\begin{eqnarray}
{d \Gamma \over d M} [B \to  D^0 \bar D^{*0} K] \approx
|c_1(\Lambda_\pi)|^2 
{\lambda^{3/2}(m_B,m_X,m_K) \over 256 \pi^3 m_B^3 m_X^2} \, q.
\label{dGam-q}
\end{eqnarray} 
%
A crude model of the crossover from the phase space distribution 
in Eq.~(\ref{dGam-q})
to the resonant distribution in Eq.~(\ref{dGam}) is a sudden 
but continuous transition at $q=\Lambda_\pi$, as illustrated in 
Figure~\ref{fig:invmass}.  This requires 
\begin{eqnarray}
|c_1(\Lambda_\pi)| \approx  {2 |\hat c_1 + \hat c_2| \over \pi \Lambda_\pi}.
\label{c1-approx}
\end{eqnarray} 
%
The integral of $d\Gamma/dM$ over the region $0<q< \Lambda_\pi$ 
increases with $a$.
In the limit $a \to \infty$, it is 3 times larger than the integral 
of a phase space distribution normalized to the same value at
$q=\Lambda_\pi$.

The Babar collaboration has measured the branching fractions for 
$B \to D^0 \bar D^{*0} K$ and $B \to  D^{*0} \bar D^0 K$
using a data sample of about $8 \times 10^7$ 
$B \bar B$ events \cite{Aubert:2003jq}.
The strongest signal was observed in the channel 
$B^+ \to  D^{*0} \bar D^0 K^+$:
$221 \pm 27$ events above the background, but with a contamination 
of about 37 events due to crossfeed from other decay channels.
If the invariant mass distributions could be measured
with resolution much better than $m_\pi^2/(2 \mu_{DD^*}) \simeq 10$ MeV
and if the histograms included enough events,
one could actually resolve the resonant enhancement near threshold
that is illustrated in Figure~\ref{fig:invmass} 
and determine both $a$ and $|\hat c_1 + \hat c_2|^2$ 
directly from the data.   The resolution that would be required 
may not be out of the question, since Babar has presented a histogram 
of $d\Gamma/dM$ for the decay $B^0 \to D^{*-} \bar D^{*0} K^+$ 
with 20 MeV bins \cite{Aubert:2003jq}.
However the region $q < m_\pi$ in which the enhancement 
is expected to occur accounts for only about $0.2$\% of the 
available phase space for the decay $B \to D^0 \bar D^{*0} K$.
Even with an enhancement in this region by a factor of 3 from a very large 
scattering length, it may be difficult to accumulate enough 
events in this region to resolve the structure in Figure~\ref{fig:invmass}.

\section{The Decay $\bm{B \to X K}$}
\label{sec:XK}

\begin{figure}
\includegraphics[width=15cm]{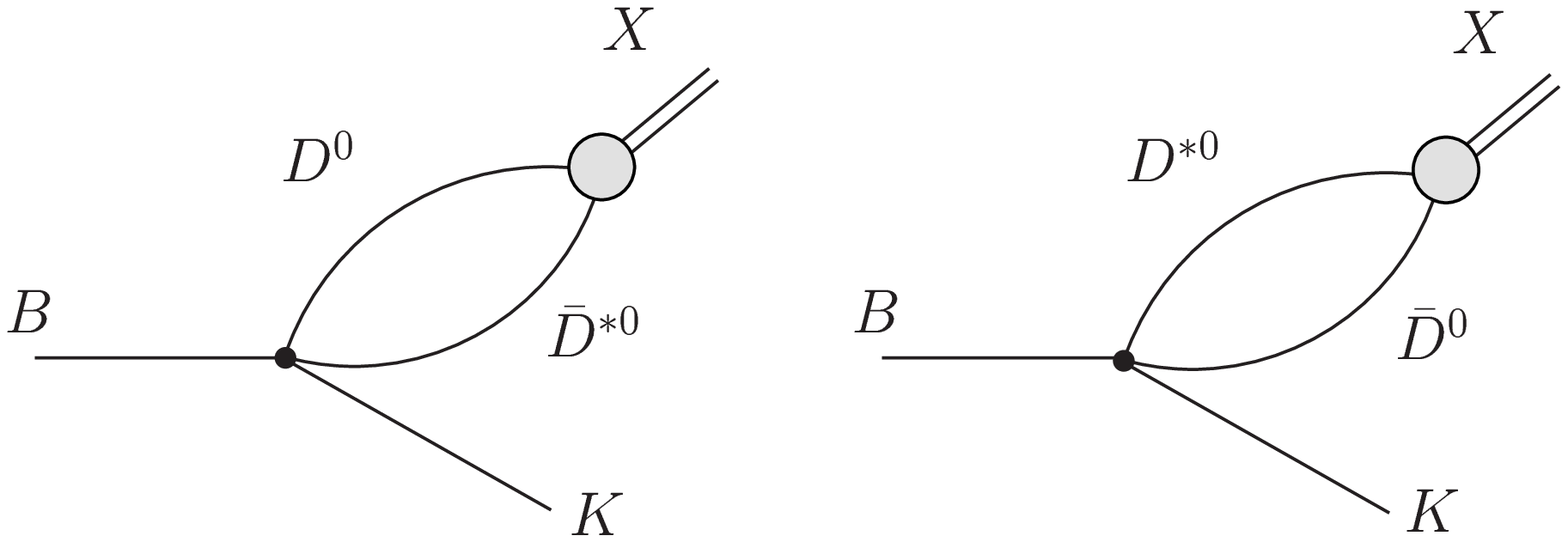}
\caption{
Feynman diagrams for the decay $B \to X K$,
with the short-distance decay amplitudes represented by dots 
and the long-distance coalesence amplitudes represented by blobs.
\label{fig:B-XK}}
\end{figure}

We proceed to apply the separation of the long-distance scale $a$
from the shorter distance scales of QCD to the decay process 
$B \to X K$.  
The amplitude for $B \to X K$ can be decomposed into two terms
corresponding to the two diagrams in Fig.~\ref{fig:B-XK}.
The first diagram represents the decay 
$B\to D^0 \bar D^{*0} K$  at short distances followed by the 
coalescence of $D^0 \bar D^{*0}$ into $X$ at long distances.
The second diagram represents the decay 
$B\to D^{*0} \bar D^0 K$ at short distances followed by the 
coalescence of $D^{*0} \bar D^0$ into $X$ at long distances.
We denote the 4-momenta of the $B$, $X$, and $K$
by $P$, $Q$, and $k$, respectively.  The expression for the amplitude is
\begin{eqnarray}
{\cal A}[B\to XK] &=& 
- i  \int \! \! \frac{d^4\ell}{(2\pi)^4} \, 
{\cal A}_{\rm short}[B\to  D^0 \bar D^{*0} K] 
\nonumber
\\
&& \hspace{1.5cm} \times 
D(q+\ell,m_{D^0}) \, D(q_*-\ell,m_{D^{*0}}) \, 
{\cal A}[ D^0 \bar D^{*0}\to X ]
\nonumber
\\
&&  - i  \int \! \! \frac{d^4\ell}{(2\pi)^4} \, 
{\cal A}_{\rm short}[B\to  D^{*0} \bar D^0 K] 
\nonumber
\\
&& \hspace{1.5cm} \times 
D(q+\ell,m_{D^0}) \, D(q_*-\ell,m_{D^{*0}}) \, 
{\cal A}[ D^{*0} \bar D^0 \to X ],
\label{ampX-sum}
\end{eqnarray}
%
where $q^\mu = (m_{D^0}/m_X) Q^\mu$ 
and $q_*^\mu = (m_{D^{*0}}/m_X) Q^\mu$
are 4-momenta that add up to the 4-momentum $Q^\mu$ of the $X$. 

\begin{figure}
\includegraphics[width=15cm]{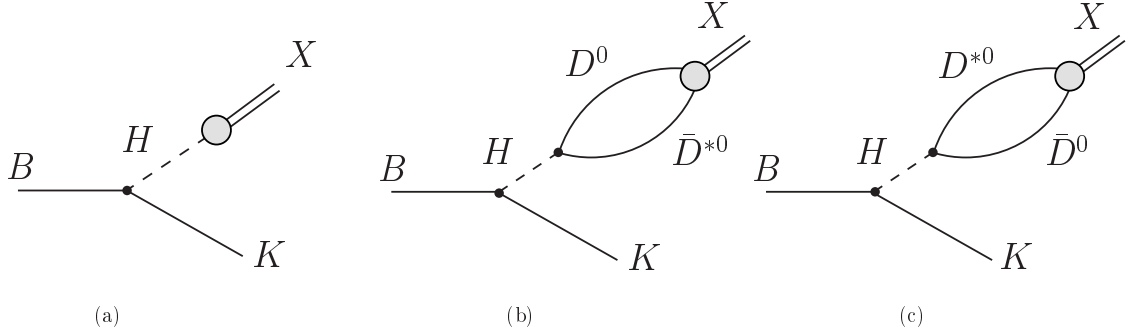}
\caption{
Feynman diagrams for the decay $B \to X K$:
(a) the diagram corresponding to the production of the hadron $H$ 
at short distances, with the blob representing the 
probability factor $Z_H^{1/2}$,
(b)--(c) equivalent diagrams obtained by iterating the 
bound-state equation to get $D D^*$ components of the 
wavefunction.
\label{fig:B-XKshort}}
\end{figure}

One might ask why we do not include the diagram in 
Fig.~\ref{fig:B-XKshort}(a), which represents the direct production 
of $X$ through the decay $B \to H K$ at short distances,
where $H$ is one of the hadronic states that appears in the
schematic decomposition of the wavefunction of $X$ in Eq.~(\ref{X}).
Such a short-distance term could be expressed in the form
$Z_H^{1/2} {\cal A}_{\rm short}[B\to H K]$ 
and can be interpreted as a contribution from a ``core component'' 
of the $X$ \cite{Voloshin:2004mh}.   
The reason such a diagram need not be considered is that it is 
already taken into account through the diagrams in Fig.~\ref{fig:B-XK}.
The various components of the wavefunction for the bound state 
$X$ satisfy coupled integral equations. 
By iterating the integral equations, one can always eliminate
the component $H$ in terms of a $D D^*$ component.
Thus the diagram in Fig.~\ref{fig:B-XKshort}(a)
can be expressed in terms of diagrams 
with explicit $D D^*$ states as in
Figs.~\ref{fig:B-XKshort}(b) and (c).  In the separation 
of the amplitude into short-distance parts and long-distance
parts represented by Fig.~\ref{fig:B-XK}, the propagators for $H$
in Figs.~\ref{fig:B-XKshort}(b) and (c) are absorbed into the 
short-distance decay amplitude.

The long-distance coalescence amplitudes in Eq.~(\ref{ampX-sum})
are given by the universal expressions in Eqs.~(\ref{amp-coal}).
The short-distance amplitudes for the decays into $D D^* K$
are given in Eqs.~(\ref{amp-DDK}).
The loop integrals in Eq.~(\ref{ampX-sum}) can be evaluated 
as in Eqs.~(\ref{int-ell0}) and (\ref{int-psi}).  
If we keep only the leading terms for $a \Lambda \gg 1$,
the amplitude has a factor $\Lambda [ c_1(\Lambda) + c_2(\Lambda)]$.
In Section~\ref{sec:DDK}, we deduced the scaling behavior
of the coefficients $c_1$ and $c_2$ as $\Lambda$ increases:
$c_1 \to \hat c_1/\Lambda$ and
$c_2 \to \hat c_2/\Lambda$, where $\hat c_1$ and $\hat c_2$
are fixed-point coefficients.
The amplitude in Eq.~(\ref{ampX-sum}) therefore reduces to
\begin{eqnarray}
{\cal A}[B\to XK] &=& 
(\hat c_1 + \hat c_2) 
(\pi^3 \mu_{DD^*} a)^{-1/2} 
P \cdot \epsilon^*.
\label{amp1-uni}
\end{eqnarray} 
%

The resulting expression for the decay rate is
\begin{eqnarray}
\Gamma[B \to X K]
= \left| \hat c_1 + \hat c_2 \right|^2
{\lambda^{3/2}(m_B,m_X,m_K) \over 64\pi^4 m_B^3 m_X^2 \mu_{DD^*} a} .
\label{B+decay}
\end{eqnarray}
%
This formula applies equally well to the decays 
$B^+ \to X K^+$ and $B^0 \to X K^0$, with the only difference 
being the values of the fixed-point coefficients
$\hat c_1$ and $\hat c_2$.
The only sensitivity to long distances is through the factor $1/a$.
We can use the expression for the decay rate 
in Eq.~(\ref{B+decay}) to eliminate
the fixed-point coefficients from the expression for the differential
decay rate in Eq.~(\ref{dGam}):
\begin{eqnarray}
{d \Gamma \over d M} [B \to  D^0 \bar D^{*0} K] =
\Gamma[B \to  X K] \,
{\mu_{DD^*} a^3 q \over \pi (1 + a^2 q^2)} .
\label{dGam:ratio}
\end{eqnarray} 
%

\section{Analysis of $\bm{B \to {\bar D}{}^{(*)} D^{(*)} K}$ Branching Fractions}
\label{sec:Babar}

A prediction of the branching fraction for $B \to X K$ 
requires the determination of the prefactor $|\hat c_1 + \hat c_2|^2$
in the expression for the decay rate in Eq.~(\ref{B+decay}).
That same prefactor appears in the differential decay rate 
$d \Gamma/dM$ in Eq.~(\ref{dGam}) for 
$B \to D^0 \bar D^{*0} K$ in the resonant region.
Thus measurements of the $D D^*$ invariant mass distribution 
in the resonant region could in principle be used 
to predict the decay rate for $B \to X K$. 
However the resonance region $q < m_\pi$ accounts for only about 
0.2\% of the available phase space for the decay $B \to D D^* K$.
It might therefore be difficult to 
accumulate enough events to determine $|\hat c_1 + \hat c_2|^2$
directly from the data.  There is a crossover 
from the resonant distribution in Eq.~(\ref{dGam})
to the phase space distribution in Eq.~(\ref{dGam-q}) at an unknown 
momentum scale $\Lambda_\pi$.  
The fraction of the phase space in which $d \Gamma/dM$ 
is described by Eq.~(\ref{dGam-q}) should be much larger 
than the 0.2\% that corresponds to the resonant region.
If one could determine the prefactor
$|c_1(\Lambda_\pi)|^2$ in Eq.~(\ref{dGam-q}) from measurements 
of the $D D^*$ invariant mass distribution, one could then
estimate the desired factor $|\hat c_1 + \hat c_2|^2$
from the relation in Eq.~(\ref{c1-approx}), which is based on a 
crude model for the crossover.  The estimate will involve the 
unknown scale $\Lambda_\pi$, which is expected to be comparable 
to $m_\pi$.

Measurements of $d \Gamma/dM$ for the decays $B \to D D^* K$
are not available.
The Babar collaboration has however measured the branching fractions 
for decays of $B^+$ (and $B^-$) and of $B^0$ (and $\bar B^0$)
into $\bar D^{(*)} D^{(*)} K$, where $D^{(*)}$ stands for 
$D^0$, $D^+$, $D^{*0}$, or $D^{*+}$ \cite{Aubert:2003jq}. 
The branching fractions are given in 
Tables~\ref{Table-ddk+} and \ref{Table-ddk0}.
A substantial fraction of $b\to c \bar c s$ decays results 
in $\bar D^{(*)} D^{(*)} K$ final states
as predicted in Ref.~\cite{Buchalla:1995kh}.  
The sum of the branching fractions is 
($3.5\pm 0.3\pm 0.5$)\% for $B^+$ 
and ($4.3\pm 0.3\pm 0.6$)\% for $B^0$. 
An isospin analysis of these decays has been carried out 
\cite{Zito:2004kz}.
We will use this data to make a rough determination of the 
prefactor $|c_1(\Lambda_\pi)|^2$ in Eq.~(\ref{dGam-q}). 

\begin{table}[t]
\caption{Branching fractions (in \%) for $B^+ \to \bar D^{(*)}D^{(*)}K$:
measurements from Ref.~\cite{Aubert:2003jq}, our 3-parameter fit,
and our 7-parameter fit.}
\begin{center}
\begin{tabular}{l | c c c}
$B^+$ decay mode & Br [\%] & ~3-parameter fit~ & ~7-parameter fit~ \\ 
\hline
$ \bar D^{0} D^{+} K^{0}$ & $~0.18\pm 0.07\pm 0.04~$&$0.17$&$0.18$ \\ 
$ \bar D^{*0} D^{+} K^{0}$ & $0.41\pm 0.15\pm0.08$&$0.31$&$0.31$ \\
$ \bar D^{0} D^{*+} K^{0}$ & $0.52\pm 0.10\pm0.07$&$0.44$&$0.45$ \\ 
$ \bar D^{*0} D^{*+} K^{0}$ &$0.78\pm 0.23\pm0.14$&$0.86$&$0.88$ \\ 
\hline
$ \bar D^{0} D^{0} K^{+} $ &$0.19\pm 0.03\pm 0.03$&$0.17$&$0.18$ \\ 
$ \bar D^{*0} D^{0} K^{+}$ &$0.18\pm 0.07\pm 0.04$&$0.31$&$0.31$ \\ 
$ \bar D^{0} D^{*0} K^{+} $ &$0.47\pm 0.07\pm 0.07$&$0.44$&$0.50$ \\ 
$ \bar D^{*0} D^{*0} K^{+}$ &$0.53\pm 0.11\pm 0.12$& $0.86$ & $0.72$ \\
\hline
$ D^{-} D^{+} K^{+}$ &$0.00\pm 0.03\pm 0.01$ &$0$ & $0.00$\\ 
$ D^{-} D^{*+} K^{+}$ &$0.02\pm 0.02\pm 0.01$ &$0$ & $0.03$ \\ 
$ D^{*-} D^{+} K^{+}$ &$0.15\pm 0.03\pm 0.02$ &$0$ & $0.03$ \\ 
$ D^{*-} D^{*+} K^{+}$ &$0.09\pm 0.04\pm 0.02$ &$0$ &$0.13$ \\ 
\end{tabular}
\end{center}
\label{Table-ddk+}
\end{table}

\begin{table}[t]
\caption{Branching fractions (in \%) for $B^0\to \bar D^{(*)}D^{(*)}K$: 
measurements from Ref.~\cite{Aubert:2003jq}, our 3-parameter fit,
and our 7-parameter fit.
}
\begin{center}
\begin{tabular}{l | c c c}
$B^0$ decay mode& Br [\%]& ~3-parameter fit~ & ~7-parameter fit~ \\ 
\hline
$D^{-} D^{0} K^{+}$   &~$0.17\pm 0.03\pm 0.03$~& $0.16$ & $0.16$ \\ 
$D^{-} D^{*0} K^{+}$  & $0.46\pm 0.07\pm 0.07$ & $0.41$ & $0.42$ \\ 
$D^{*-} D^{0} K^{+}$  & $0.31\pm 0.04\pm 0.04$ & $0.29$ & $0.29$\\ 
$D^{*-} D^{*0} K^{+}$ & $1.18\pm 0.10\pm 0.17$ & $0.79$ & $0.81$ \\ 
\hline
$D^{-} D^{+} K^{0}$   & $0.08\pm 0.06\pm 0.03$ & $0.16$ & $0.16$ \\ 
$D^{*-} D^{+} K^{0}$, $D^{-} D^{*+} K^{0}$ 
                     & $0.65\pm 0.12\pm 0.10$ & $0.29+0.41$ & $0.29+0.46$ \\ 
$D^{*-} D^{*+} K^{0}$ & $0.88\pm 0.15\pm 0.13$ & $0.79$ & $0.67$ \\ 
\hline
$\bar D^{0} D^{0} K^{0}$   & $0.08\pm 0.04\pm 0.02$ & $0$ & $0.00$ \\ 
$\bar D^{0} D^{*0} K^{0}$, $\bar D^{*0} D^{0} K^{0}$
                           & $0.17\pm 0.14\pm0.07$ & $0+0$ & $0.02+0.02$ \\ 
$\bar D^{*0} D^{*0} K^{0}$ &$0.33\pm 0.21\pm 0.14$ & $0$  &$0.12$ \\ 
\end{tabular}
\end{center}
\label{Table-ddk0}
\end{table}

The most important terms in the 
effective weak Hamiltonian for $b\to c \bar c s$ decays 
at a renormalization scale of order $m_b$ is
\begin{eqnarray}
{\cal H}_{W}= \frac{G_{F}}{\sqrt{2}} V_{cb} V_{cs}^* 
\left( C_1 \, {\cal O}_1 + C_2 \, {\cal O}_2 \right)  + {\rm h.c.},
\end{eqnarray}
%
where $C_1$ and $C_2$ are Wilson coefficients 
and ${\cal O}_1$ and ${\cal O}_2$ are local 
four-fermion operators:
\begin{subequations}
\begin{eqnarray}
{\cal O}_1 &=& \bar c \gamma_L^\mu b \;
	\bar s {\gamma_L^{\phantom \mu}}_\mu c ,
\\
{\cal O}_2 &=& \bar s \gamma_L^\mu b \;
	\bar c {\gamma_L^{\phantom \mu}}_\mu c. 
\end{eqnarray}
\label{4fermion}
\end{subequations}
%
We have used the notation $\gamma_L^\mu = \gamma^\mu(1-\gamma_5)$.
Both operators are products of color-singlet currents.
We make the simplifying assumption that matrix elements of the
operators ${\cal O}_1$ and ${\cal O}_2$ between 
the initial-state $B$ and the final-state   
$\bar D^{(*)} D^{(*)} K$ can be factorized into 
products of matrix elements of currents.
For example, the matrix elements for decays into 
$D^0 \bar D^{*0} \bar K$ and $D^{*0} \bar D^0 \bar K$ are 
\begin{subequations}
\begin{eqnarray}
\langle D^0 \bar D^{*0} K^- | {\cal H}_W | B^- \rangle &=& 
(G_F/\sqrt{2}) V_{cb}^* V_{sc} 
\left( C_1 \,
\langle D^0 | \bar c \gamma_L^\mu  b | B^- \rangle \,
\langle \bar D^{*0} K^- | \bar s {\gamma_L^{\phantom \mu}}_\mu c | \emptyset \rangle
\right.
\nonumber 
\\
&& \left. \hspace{3cm} + \, C_2 \,  
\langle K^- | \bar s \gamma_L^\mu  b | B^- \rangle \,
\langle D^0 \bar D^{*0} | \bar c {\gamma_L^{\phantom \mu}}_\mu c | \emptyset \rangle
\right),
\label{b+dd*k}
\\
\langle D^{*0} \bar D^0 K^- | {\cal H}_W | B^- \rangle &=&
(G_F/\sqrt{2}) V_{cb}^* V_{sc} 
\left( C_1 \,
\langle D^{*0} | \bar c \gamma_L^\mu  b | B^- \rangle \,
\langle \bar D^0 K^- | \bar s {\gamma_L^{\phantom \mu}}_\mu c | \emptyset \rangle
\right.
\nonumber 
\\
&& \left. \hspace{3cm} + \, C_2 \,  
\langle K^- | \bar s \gamma_L^\mu b | B^- \rangle \,
\langle D^{*0} \bar D^0 | \bar c {\gamma_L^{\phantom \mu}}_\mu c | \emptyset \rangle
\right),
\label{b+d*dk}
\\
\langle D^0 \bar D^{*0} \bar K^0 | {\cal H}_W | \bar B^0 \rangle &=&
(G_F/\sqrt{2}) V_{cb}^* V_{sc} \, C_2 \, 
\langle \bar K^0 | \bar s \gamma_L^\mu  b | \bar B^0 \rangle \,
\langle D^0 \bar D^{*0} | \bar c {\gamma_L^{\phantom \mu}}_\mu c | \emptyset \rangle,
\label{b0dd*k}
\\
\langle D^{*0} \bar D^0 \bar K^0 | {\cal H}_W | \bar B^0 \rangle &=&
(G_F/\sqrt{2}) V_{cb}^* V_{sc} \, C_2 \,
\langle \bar K^0 | \bar s \gamma_L^\mu  b | \bar B^0 \rangle \,
\langle D^{*0} \bar D^0 | \bar c {\gamma_L^{\phantom \mu}}_\mu c | \emptyset \rangle.
\label{b0d*dk}
\end{eqnarray}
\label{BtoDDK}
\end{subequations}
%
The accuracy of the factorization assumption for this process 
has been discussed in detail in Ref.~\cite{Bauer:2002sh}.

The terms in Eqs.~(\ref{BtoDDK}) with coefficient $C_2$ are
called ``color-suppressed'' amplitudes, because $C_2$ 
is suppressed by $1/N_c$ relative to $C_1$.  Only the color-suppressed 
amplitudes contribute to the decays $B^+ \to \bar D^{(*)}D^{(*)}K^+$
with $\bar D^{(*)}$ and $D^{(*)}$ both charged and to the decays 
$B^0 \to \bar D^{(*)}D^{(*)}K^0$ with $\bar D^{(*)}$ and $D^{(*)}$ both neutral.  
In particular, the only contributions to the decays of 
$B^0$ into $D^0 \bar D^{*0} K^0$ and $D^{*0} \bar D^0 K^0$ 
are from the color-suppressed amplitudes.
As is evident in 
Tables~\ref{Table-ddk+} and \ref{Table-ddk0}, the branching fractions 
for these color-suppressed channels are observed to be significantly 
smaller than those for other decay channels. 

Lorentz invariance can be used to reduce each of the current matrix
elements to a linear combination of independent tensor structures 
whose coefficients are form factors.  
Heavy quark symmetry provides constraints 
between the form factors that can be deduced using the covariant 
representation formalism decribed in Ref.~\cite{Manohar-Wise}.
Matrix elements of operators with a heavy quark field 
$Q$ (or $\bar Q$) and a $Q \bar q$ meson in the initial 
(or final) state can be expressed in terms of a heavy meson field 
$H_v$ (or $\bar H_v$) defined by
\begin{subequations}
\begin{eqnarray}
H_v &=& \frac{1+\fsl{v}}{2}
\left[ V_v^\mu \gamma_\mu + i P_v \gamma_5 \right],
\\
\bar H_v &=& 
\left[ V_v^{\mu \dagger} \gamma_\mu + i P_v^\dagger \gamma_5 \right] 
\frac{1+\fsl{v}}{2},
\end{eqnarray}
\end{subequations}
%
where $V_v^\mu$ and $P_v$ are operators that
annihilate vector and pseudoscalar $Q \bar q$ mesons with 4-velocity $v$.
We also require the matrix elements of operators with
a heavy quark field $Q$  and a $\bar Q q$ meson 
in the final state.  They can be expressed in terms of a 
heavy meson field $H'_v$ that creates 
a $\bar Q q$ meson with 4-velocity $v$:
\begin{eqnarray}
H'_v &=& \frac{1-\fsl{v}}{2}
\left[ V_v^\mu \gamma_\mu- i P_v\gamma_5 \right],
\end{eqnarray}
%
The relative phase between the $V_v^\mu$ and $P_v$ terms
has been deduced by demanding that vacuum-to-$D^{(*)}\bar D^{(*)}$ 
matrix elements of operators of the form $\bar Q \Gamma Q$
have the correct charge conjugation properties.

We now list the expressions for the matrix elements of the
currents that follow from heavy-quark symmetry.
We denote the velocity 4-vectors of the $\bar B$, $D^{(*)}$,
and $\bar D^{(*)}$ by $V$, $v$ and $\bar v$, respectively.
We denote the polarization 4-vectors of the $D^*$ 
and $\bar D^*$ by $\epsilon$ and $\bar \epsilon$, respectively.
They satisfy $v \cdot \epsilon = 0$ 
and $\bar v \cdot \bar \epsilon = 0$.
The $\bar B$-to-$D^{(*)}$ matrix elements are
\begin{subequations}
\begin{eqnarray}
\langle D(v) | \bar c \gamma_L^\mu b | \bar B(V) \rangle 
&=&  
\xi(w)(v+V)^\mu ,
\\
\langle D^{*}(v,\epsilon) | \bar c \gamma_L^\mu b | \bar B(V) \rangle 
&=& 
i\xi(w)\left[ (1 + v \cdot V) \epsilon^\mu - (V \cdot \epsilon) v^\mu 
-i \varepsilon^\mu(v,V,\epsilon) \right],
\end{eqnarray}
\label{BtoD}
\end{subequations} 
%
where the form factor $\xi$ is a function of $w=v\cdot V$.
We have used the notation
$\varepsilon^\mu(p,q,r) 
= \varepsilon^{\mu\nu\alpha\beta} p_\nu q_\alpha r_\beta$
and the sign convention $\varepsilon^{0123} = +1$.
The vacuum-to-$\bar D^{(*)} \bar K$ matrix elements are
\begin{subequations}
\begin{eqnarray}
\langle \bar D(\bar v) \bar K(k) | \bar s \gamma_L^\mu c | \emptyset \rangle
&=& \eta_1(\kappa) \, \bar v^\mu + \eta_2(\kappa) \, k^\mu,
\\
\langle \bar D^*(\bar v,\bar \epsilon) \bar K(k) | \bar s \gamma_L^\mu c 
	| \emptyset \rangle
&=&  -i\eta_1(\kappa) \, \bar \epsilon^\mu
-i\eta_2(\kappa)\, 
\left[ (\bar v \cdot k) \bar \epsilon^\mu - (k \cdot \bar \epsilon) \bar v^\mu 
	+ i \varepsilon^\mu(\bar v,k,\bar \epsilon) \right],
\nonumber
\\
&& \label{eq:DK}
\end{eqnarray}
\label{0toDK}
\end{subequations}
%
where the form factors $\eta_1$ and $\eta_2$ 
are functions of $\kappa = \bar v \cdot k$.
The vacuum-to-$D^{(*)}\bar D^{(*)}$ matrix elements are
\begin{subequations}
\begin{eqnarray} 
\langle D(v) \bar D(\bar v) | \bar c \gamma_L^\mu c | \emptyset \rangle 
&=&  
\zeta(w')(v-\bar v)^\mu,
\label{eq:DD-1}
\\
\langle D(v) \bar D^*(\bar v,\bar \epsilon) | \bar c \gamma_L^\mu c
	| \emptyset \rangle 
&=& 
i\zeta(w') 
\left[ (1-v\cdot \bar v) \bar \epsilon^\mu + (v \cdot \bar \epsilon) \bar v^\mu 
	+i \varepsilon^\mu(v,\bar v,\bar \epsilon) \right],
\label{eq:DD-2}
\\
\langle D^* (v,\epsilon) \bar D(\bar v) | \bar c \gamma_L^\mu c 
	| \emptyset \rangle 
&=& 
i\zeta(w')
\left[ (1-v\cdot \bar v) \epsilon^\mu + (\bar v \cdot \epsilon) v^\mu
	+i \varepsilon^\mu(v,\bar v,\epsilon) \right],
\label{eq:DD-3}
\\
\langle D^*(v,\epsilon) \bar D^*(\bar v,\bar \epsilon) |
	\bar c \gamma_L^\mu c | \emptyset \rangle 
&=& 
\zeta(w') 
\left[ \epsilon \cdot \bar \epsilon (v - \bar v)^\mu
	+ (\bar v \cdot \epsilon) \bar \epsilon^\mu
	- (v \cdot \bar \epsilon) \epsilon^\mu 
	- i \varepsilon^\mu(v-\bar v,\epsilon,\bar \epsilon) \right],
\nonumber 
\\ 
&& 
\label{eq:DD-4}
\end{eqnarray}
\label{0toDD}
\end{subequations}
%
where the form factor $\zeta$ is a function of $w'=v\cdot \bar v$.
The $\bar B$-to-$\bar K$ matrix elements are  
\begin{eqnarray}
\langle \bar K(k)| \bar s \gamma_L^\mu b | \bar B (V) \rangle 
=\omega_1(\kappa') \, V^\mu + \omega_2(\kappa') \, k^\mu,
\label{BtoK}
\end{eqnarray}
%
where the form factors $\omega_1$ and $\omega_2$ 
are functions of $\kappa' = V \cdot k$.  
In the current matrix elements in Eqs.~(\ref{BtoD}), 
(\ref{0toDK}), (\ref{0toDD}), and (\ref{BtoK}),
the heavy meson states have the standard nonrelativistic
normalizations. To obtain the standard relativistic normalizations,
matrix elements involving $B$, $D$ or $\bar D$, and $D^*$ or $\bar D^*$
must be multiplied by $m_B^{1/2}$, $m_D^{1/2}$,
and $m_{D^*}^{1/2}$, respectively.

The amplitudes for the decays $\bar B \to \bar D^{(*)} D^{(*)} \bar K$ 
at leading-order in $\Lambda_{\rm QCD}/m_b$ and $\Lambda_{\rm QCD}/m_c$ 
are obtained by inserting the current matrix elements
in Eqs.~(\ref{BtoD}), (\ref{0toDK}), (\ref{0toDD}), and (\ref{BtoK}) 
into the factorized expressions for the decay amplitudes,
such as those in Eqs.~(\ref{BtoDDK}). 
For example, the amplitudes for the decays into 
$D^0 \bar D^{*0} \bar K$ and $D^{*0} \bar D^0 \bar K$ are
\begin{subequations}
\begin{eqnarray}
{\cal A}[B^- \to D^0 \bar D^{*0} K^-] &=& 
-i G_1 (v + V) \cdot \epsilon 
\nonumber 
\\ 
&& 
-i (G_2/m_B)
\left[ v_* \cdot k \, (v + V) \cdot \epsilon
	- v_* \cdot (v + V) \, k \cdot \epsilon
	+ i \varepsilon(v+V,v_*,k,\epsilon) \right]
\nonumber 
\\
&& 
+i G_3
\left[ (1 - v \cdot v_*) \, V \cdot \epsilon + (v_* \cdot V) \, v \cdot \epsilon
	+i \varepsilon(v,v_*,V,\epsilon) \right]
\nonumber 
\\
&& 
+i (G_4/m_B)
\left[ (1 - v \cdot v_*) \, k \cdot \epsilon + (v_* \cdot k) \, v \cdot \epsilon
	+i \varepsilon(v,v_*,k,\epsilon)  \right],
\label{A-Dbar*D}
\\
{\cal A}[B^- \to D^{*0} \bar D^0 K^-] &=& 
i G_1 
\left[ (1+ v_* \cdot V) \, v \cdot \epsilon - (v \cdot v_*) \, V \cdot \epsilon
	- i \varepsilon(v,v_*,V,\epsilon) \right] 
\nonumber 
\\
&& 
+i (G_2/m_B) 
\left[ (1+ v_* \cdot V) \, k \cdot \epsilon  - (v_* \cdot k) \, V \cdot \epsilon
	- i \varepsilon(v_*,V,k,\epsilon) \right] 
\nonumber 
\\
&&
+i G_3 
\left[ (1 - v \cdot v_*) \, V \cdot \epsilon
	+ (v_* \cdot V) \, v \cdot \epsilon
	-i \varepsilon(v,v_*,V,\epsilon) \right]
\nonumber 
\\
&& 
+i (G_4/m_B)
\left[ (1 - v \cdot v_*) \, k \cdot \epsilon
	+ (v_* \cdot k) \, v \cdot \epsilon
	-i \varepsilon(v,v_*,k,\epsilon) \right],
\label{A-DbarD*}
\\
{\cal A}[\bar B^0 \to D^0 \bar D^{*0} \bar K^0] &=& 
i G_3 
\left[ (1- v \cdot v_*) \, V \cdot \epsilon
	+ (v_* \cdot V) \, v \cdot \epsilon
	+i \varepsilon(v,v_*,V,\epsilon) \right]
\nonumber 
\\ 
&& 
+i (G_4/m_B) 
\left[ (1- v \cdot v_*) \, k \cdot \epsilon 
	+ (v_* \cdot k) \, v \cdot \epsilon
	+i \varepsilon(v,v_*,k,\epsilon) \right],
\label{A0-Dbar*D}
\\
{\cal A}[\bar B^0 \to D^{*0} \bar D^0 \bar K^0] &=& 
i G_3 
\left[ (1- v \cdot v_*) \, V \cdot \epsilon
	+ (v_* \cdot V) \, v \cdot \epsilon
	-i \varepsilon(v,v_*,V,\epsilon) \right]
\nonumber 
\\ 
&&
+i (G_4/m_B) 
\left[ (1- v \cdot v_*) k \cdot \epsilon 
	+ (v_* \cdot k) v \cdot \epsilon
	-i \varepsilon(v,v_*,k,\epsilon) \right],
\label{A0-DbarD*}
\end{eqnarray}
\end{subequations}
%
where $V$, $v$, and $v_*$ are the velocity 4-vectors of the
$B$, $D$, and $D^*$ and $\epsilon$ is the polarization 4-vector
of the $D^*$ which satisfies $v_* \cdot \epsilon = 0$.
We have used the notation 
$\varepsilon(p,q,r,s)
= \varepsilon^{\mu\nu\alpha\beta} p_\mu q_\nu r_\alpha s_\beta$.
The four independent dimensionless 
form factors are
\begin{subequations}
\begin{eqnarray}
G_1((P-q)^2) &=& (G_F/\sqrt{2}) V_{cb}^* V_{sc} \, C_1 \, 
( m_B m_{D^0} m_{D^{*0}} )^{1/2} \, 
\xi(v \cdot V) \, \eta_1(v_* \cdot k),
\\
G_2((P-q_*)^2) &=& (G_F/\sqrt{2}) V_{cb}^* V_{sc} \, C_1 \,
( m_B^3 m_{D^0} m_{D^{*0}} )^{1/2} \, 
\xi(v_* \cdot V) \, \eta_2(v \cdot k),
\\
G_3((P-k)^2) &=& (G_F/\sqrt{2}) V_{cb}^* V_{sc} \, C_2 \,
( m_B m_{D^0} m_{D^{*0}} )^{1/2} \, 
\zeta(v \cdot v_*) \, \omega_1(V \cdot k),
\\
G_4((P-k)^2) &=& (G_F/\sqrt{2}) V_{cb}^* V_{sc} \, C_2 \,
( m_B^3 m_{D^0} m_{D^{*0}} )^{1/2} \, 
\zeta(v \cdot v_*) \, \omega_2(V \cdot k).
\end{eqnarray} 
\end{subequations}
%
The amplitudes for the other $\bar B \to \bar D^{(*)} D^{(*)} \bar K$ 
decays are obtained similarly. 
Isospin symmetry, in addition to the factorization assumption 
and heavy quark symmetry, can be used to express all 24 decay 
amplitudes in terms of the four form factors
$G_1(q^2)$, $G_2(q^2)$, $G_3(q^2)$, and $G_4(q^2)$.

We proceed to use our expressions for the decay amplitudes
to analyze the data from the Babar collaboration on the branching 
fractions for $B \to\bar D^{(*)} D^{(*)} K$ \cite{Aubert:2003jq}. 
For simplicity, we approximate the form factors $G_{i}(q^2)$ by constants. 
We can choose the overall phase so that $G_1$ is real-valued.
After integrating over the phase space, we obtain expressions 
for the branching fractions that are quadratic in the constants 
$G_i$ and their complex conjugates.  
The Babar data consists of the 12 branching fractions for $B^+$ 
given in Table~\ref{Table-ddk+} and the 
10 branching fractions for $B^0$ given in Table~\ref{Table-ddk0}.   
For each of the data points, we add the statistical and systematic 
errors in quadrature.  We then determine the best fits for the 
constants $G_i$ by minimizing the $\chi^2$ for the 22 data points.

The decays $B^+ \to \bar D^{(*)} D^{(*)} K^+$
with $\bar D^{(*)}$ and $D^{(*)}$ both charged and 
$B^0 \to \bar D^{(*)} D^{(*)} K^0$ with $\bar D^{(*)}$ and $D^{(*)}$ 
both neutral have branching fractions that are
significantly smaller than other decay channels.  The only factorizable
contributions to their decay amplitudes come from the color-suppressed 
amplitudes with form factors $G_3$ and $G_4$.  Their small 
branching fractions motivates a simplified analysis in which
$G_3$ and $G_4$ are set to 0.  The only parameters that remain
are the real constant $G_1$ and the complex constant $G_2$.  
Thus there are 3 real parameters to fit the 22 branching fractions. 
The parameters that minimize the $\chi^2$  are
\begin{subequations}
\begin{eqnarray}
  G_1 &=& 1.9 \times 10^{-6}, 
\\
  G_2 &=& (-21.2 + 5.5i)\times 10^{-6}. 
\end{eqnarray}
\label{G12}
\end{subequations}
%
The fitted value of $G_2$ is about an order of magnitude larger 
than that of $G_1$.
The branching fractions for this 3-parameter fit 
are shown in Tables~\ref{Table-ddk+} and \ref{Table-ddk0}. 
The $\chi^2$ per degree of freedom is $42.0/19 = 2.2$.
There are 7 decay modes for which the deviations from the data are 
significantly larger than one standard deviation,
including $B^+ \to \bar D^{*0} D^0 K^+$.

We have also carried out a fit that allows nonzero 
values of the color-suppressed form factors $G_3$ and $G_4$.
If these form factors are approximated by complex-valued constants,
there are 7 real parameters to fit the
22 branching fractions. 
The parameters that minimize the $\chi^2$  are
\begin{subequations}
\begin{eqnarray}
  G_1 &=& 1.8 \times 10^{-6}, 
\\
  G_2 &=& (-21.6 + 5.0i)\times 10^{-6}, 
\\
  G_3 &=& (2.6+0.01i)\times 10^{-6}, 
\\
  G_4 &=& (-1.5-0.7i)\times 10^{-6}. 
\end{eqnarray}
\label{G1234}
\end{subequations}
%
Note that the values of $G_1$ and $G_2$ are essentially identical
to those from the 3-parameter fit in Eqs.~(\ref{G12}).
The branching fractions for this 7-parameter fit 
are shown in Tables~\ref{Table-ddk+} and \ref{Table-ddk0}. 
The $\chi^2$ per degree of freedom is $29.2/15 = 1.9$.
There are still 4 decay modes for which the deviations from the data are 
significantly larger than one standard deviation,
including $B^+ \to \bar D^{*0} D^0 K^+$.

One could of course improve the fits to the branching fractions 
by allowing for dependence of each the form factors $G_1$, $G_2$, $G_3$,
and $G_4$ on the appropriate momentum transfer $q^2$.  
However allowing even for linear dependence on $q^2$ 
would introduce 8 additional real parameters. 
Such an analysis might be worthwhile if Dalitz plots
for the decays were available and could also be used in the fits.

\section{Predictions for $B\to  X K$ Decays}
\label{sec:predictions}

In this section, we use the results of our analysis of 
the branching fractions for $B \to D^{(*)} \bar D^{(*)} K$ 
to estimate the branching fractions for the decays
$B^+ \to X K^+$ and $B^0 \to X K^0$. 
Our strategy once again is to use that data to provide a rough 
determination of the prefactor $|c_1(\Lambda_\pi)|^2$
in the differential decay rate $d\Gamma/dM$ for $B \to D^0 \bar D^{*0} K$
in the region near the $D D^*$ threshold where the $D D^*$ 
invariant mass distribution follows the phase space distribution 
in Eq.~(\ref{dGam-q}).
The crossover to the resonant distribution in Eq.~(\ref{dGam})
occurs at an unknown momentum scale $\Lambda_\pi$,
which is expected to be comparable to $m_\pi$.
Given a value for $|c_1(\Lambda_\pi)|^2$, we can use the relation 
in Eq.~(\ref{c1-approx}), which follows from a 
crude model for the crossover, to estimate $|\hat c_1 + \hat c_2|^2$.
This value can then be inserted in Eq.~(\ref{B+decay})
to get an estimate of the decay rate for $B \to X K$.

We first consider the decay $B^+ \to X K^+$, whose branching fraction 
should be the same as for $B^- \to X K^-$.
The coefficient $c_1(\Lambda_\pi)$ for the decay 
$B^- \to  D^0 \bar D^{*0} K^-$ and the corresponding 
coefficient $c_2(\Lambda_\pi)$ for the decay 
$B^- \to  D^{*0} \bar D^0 K^-$ can be deduced by matching 
the amplitudes in Eqs.~(\ref{A-Dbar*D}) 
and (\ref{A-DbarD*}) at the $D D^*$ threshold
to the expressions in Eqs.~(\ref{amp-DDK}):
\begin{eqnarray}
c_1(\Lambda_\pi) = c_2(\Lambda_\pi) 
= -iG_1/m_B+iG_2(m_B+m_D+m_{D^*})/m_B^2.
\end{eqnarray}
%
Using the numerical values for $G_1$ and $G_2$ in 
either Eqs.~(\ref{G12}) or Eqs.~(\ref{G1234}), 
the estimate from Eq.~(\ref{c1-approx}) is
\begin{eqnarray}
|\hat c_1 + \hat c_2| \approx 1.6 \times 10^{-6} \, \Lambda_\pi/m_\pi .
\label{c12}
\end{eqnarray}
%
Inserting this into the expression for the decay rate in
Eq.~(\ref{B+decay}) and dividing by the measured width of the $B^+$,
we obtain 
\begin{eqnarray}
 {\rm Br}[B^+\to XK^+] \approx 2.7 \times 10^{-5}
 \left( \frac{\Lambda_\pi}{m_\pi} \right)^2
 \left(\frac{E_b}{0.5 \;{\rm MeV}}\right)^{1/2}.
\label{BrBXK}
\end{eqnarray}
%
Our previous analysis in Ref.~\cite{Braaten:2004fk}
used the four branching fractions for $B^+$ to decay into
$D^0 \bar D^0 K^+$,  $D^0 \bar D^{*0} K^+$,  $D^{*0} \bar D^0 K^+$,  
and $D^{*0} \bar D^{*0} K^+$ to fit the constants $G_1$ and $G_2$.
The final result was identical to Eq.~(\ref{BrBXK}) except
that the numerical value of the branching fraction for 
$\Lambda_\pi = m_\pi$ and $E_b=0.5$ MeV was $2.9 \times 10^{-5}$.
The estimate in Eq.~(\ref{BrBXK}) is sensitive to the unknown momentum
scale $\Lambda_\pi$  at which the 
invariant mass distribution crosses over from the phase space 
distribution in Eq.~(\ref{dGam-q}) to the resonant distribution 
in Eq.~(\ref{dGam}).
The natural scale for $\Lambda_\pi$ may be $m_\pi$, but we should not 
be surprised if it differs by a factor of 2 or 3.
Thus the result in Eq.~(\ref{BrBXK}) is only an order-of-magnitude 
estimate of the branching fraction. 
It can be compared to the product of the branching fractions for
$B^+ \to X K^+$ and $X \to J/\psi \, \pi^+ \pi^-$ in Eq.~(\ref{Brexp}).
Our estimate is compatible with this measurement if  
$J/\psi \, \pi^+ \pi^-$ is one of the major decay modes of $X$.
If $E_b = 0.5$ MeV and if we allow for $\Lambda_\pi$ 
to differ from $m_\pi$ by a factor of 2, the branching fraction 
for $X \to J/\psi \, \pi^+ \pi^-$ should be greater than $10^{-1}$.

We next consider the decay $B^0 \to X K^0$, whose branching fraction 
should be the same as for $\bar B^0 \to X \bar K^0$.
The amplitudes in Eqs.~(\ref{A0-Dbar*D}) and (\ref{A0-DbarD*}) approach 0
as the $D D^*$ approaches its threshold.  Thus our assumptions 
of factorization and heavy quark symmetry imply
that $c_1(\Lambda_\pi)=c_2(\Lambda_\pi)=0$ for this decay.
We proceed to consider the size of the coefficients 
that would be expected from the violation of these assumptions.
The factorization assumption for the 
$B \to \bar D^{(*)} D^{(*)} K$ amplitudes can be justified 
by the large $N_c$ limit.  Since we have included terms
up to $O(1/N_c)$ in the amplitude, we expect the deviations
from the factorization assumptions to be $O(1/N_c^2)$ in the amplitude. 
Violation of heavy quark symmetry would give rise to 
terms of $O(\Lambda_{\rm QCD}/m_c)$ in the amplitudes.
We expect the largest nonzero contributions to the coefficients 
$c_1(\Lambda_\pi)$ and $c_2(\Lambda_\pi)$ to come from 
violations of heavy quark symmetry.

To obtain an estimate of the decay rate for $B^0\to XK^0$,
we relax the assumption of heavy quark symmetry.
Lorentz invariance allows three independent tensor structures
in the matrix elements
$\langle D \bar D^* | \bar c \gamma_L^\mu c | \emptyset \rangle$
and
$\langle D^* \bar D | \bar c \gamma_L^\mu c | \emptyset \rangle$,
but heavy quark symmetry requires those terms to enter in the 
particular linear combinations given 
in Eqs.~(\ref{eq:DD-2}) and (\ref{eq:DD-3}).
Lorentz invariance implies that
only one of the three independent terms can be nonzero at the 
$D D^*$ threshold:  the $\bar \epsilon^\mu$ term in Eq.~(\ref{eq:DD-2}) 
and the $\epsilon^\mu$ term in Eq.~(\ref{eq:DD-3}).
Heavy quark symmetry constrains the coefficients 
of $\bar \epsilon^\mu$ and $\epsilon^\mu$ to be
$i \zeta(w') (1 - v \cdot \bar v)$, which vanishes at the threshold. 
The constraint of heavy quark symmetry can be relaxed by adding
to the coefficients of $\bar \epsilon^\mu$ in Eq.~(\ref{eq:DD-2})
and $\epsilon^\mu$ in Eq.~(\ref{eq:DD-3})
the term $i \chi \zeta(1)$, which is nonzero at the threshold.
This corresponds to adding the terms 
$i \chi [G_3 (V \cdot \epsilon) + G_4/m_B (k \cdot \epsilon)]$ 
to the amplitudes in Eqs.~(\ref{A0-Dbar*D}) and (\ref{A0-DbarD*}).
In Table~\ref{Table-ddk0}, the 7-parameter fit gives 0.04 
for the sum of the two branching fractions for $B^0$ to decay into
$D^0 \bar D^{*0} K^0$ and $\bar D^0 D^{*0} K^0$, which
is about one standard deviation below the measured value.
The complex parameter $\chi$ can be adjusted so that the sum 
of the two branching fractions is equal to the central value 0.17 
given in Table~\ref{Table-ddk0}.
Using the values of $G_3$ and $G_4$ in Eq.~(\ref{G1234}),
the required values of $\chi$ form a curve that passes through 
the real values $\chi = -1.9$ and $\chi = 5.8$
and the imaginary values $\chi = \pm 3.3 \, i$.
If $\chi$ is allowed to vary over the region in which the sum 
of the two branching fractions is within one standard deviation 
of the central value, its absolute value has the range 
$0 < |\chi| < 7.3$.

We proceed to make a quantitative estimate of 
the decay rate for $B^0 \to X K^0$.
The coefficient $c_1(\Lambda_\pi)$ for the decay 
$\bar B^0 \to  D^0 \bar D^{*0} \bar K^0$ and the corresponding 
coefficient $c_2(\Lambda_\pi)$ for the decay 
$\bar B^0 \to  D^{*0} \bar D^0 \bar K^0$ can be deduced by matching 
the amplitudes 
$i \chi [G_3 (V \cdot \epsilon) + G_4/m_B (k \cdot \epsilon)]$ 
to the expressions in Eqs.~(\ref{amp-DDK}):
\begin{eqnarray}
 c_1(\Lambda_\pi)=c_2(\Lambda_\pi)= i \chi (G_3+G_4)/m_B.
\end{eqnarray}
%
If we use the estimate in Eq.~(\ref{c1-approx}) to deduce the values
of $|\hat c_1 + \hat c_2|^2$
for both $B^0 \to X K^0$ and $B^+ \to X K^+$, 
the ratio of their branching fractions is
\begin{eqnarray}
{{\rm Br}[B^0 \to X K^0] \over {\rm Br}[B^+\to XK^+]}
\approx {|\chi|^2 |G_3 + G_4|^2 \over |G_1 - G_2 (m_B+m_D+m_{D^*})/m_B|^2} 
\, {\tau[B^0] \over \tau[B^+]}.
\label{Gratio}
\end{eqnarray}
The ratio of the lifetimes of the $B^0$ and $B^+$ is $0.921 \pm 0.014$.
If $\chi$ is allowed to vary over 
the region $0 < |\chi| < 7.3$, the ratio in Eq.~(\ref{Gratio})
ranges from 0 to $8 \times 10^{-2}$. 
We conclude that the branching fraction for $B^0 \to X K^0$ 
is likely to be suppressed by at least an order of magnitude 
compared to that for $B^+ \to X K^+$.

\section{Summary}
\label{sec:summary}

If the $X(3872)$ is a loosely-bound S-wave molecule corresponding
to a $C=+$ superposition of $D^0 \bar D^{*0}$ and $D^{*0} \bar D^0$, 
these charm mesons necessarily have a scattering length that is 
large compared to all other length scales of QCD.
The $X$ can be produced through the weak decay 
of the $B$ meson into $D^0 \bar D^{*0} K$ or $D^{*0} \bar D^0 K$
at short distances followed by the coalescence of the charm mesons 
at the long-distance scale $a$. 
We have analyzed the decay $B \to X K$ and the decays of $B$ into 
$D^0 \bar D^{*0} K$ and $D^{*0} \bar D^0 K$ near the threshold 
for the charm mesons by separating the decay
amplitudes into short-distance factors and long-distance factors.
The long-distance factors are determined by $a$, 
while the short-distance factors are essentially 
determined by the amplitudes for $B \to D^0 \bar D^{*0} K$ 
and $B \to D^{*0} \bar D^0 K$ at $D D^*$ invariant masses 
that are a little above the resonance region,
which extends to about 10 or 20 MeV above the threshold.
We obtained a crude determination of the short-distance amplitudes by 
analyzing data from the Babar collaboration 
on the branching fractions for $B\to \bar D^{(*)} D^{(*)} K$
using a factorization assumption, heavy quark symmetry, and 
isospin symmetry.  

Our estimate for the branching fraction for $B^+ \to X K^+$ 
is given in Eq.~(\ref{BrBXK}).  It scales with the binding energy 
$E_b$ of $X$ as $E_b^{1/2}$.  It also scales as $\Lambda_\pi^2$,
where $\Lambda_\pi$ is an unknown crossover momentum scale 
that is expected to be comparable to $m_\pi$.  
If we take $E_b = 0.5$ MeV and if we allow
$\Lambda_\pi$ to vary between $m_\pi/2$ and $2 m_\pi$,
our estimate of the branching fraction varies from about 
$7 \times 10^{-6}$ to about $1 \times 10^{-4}$.  This range is 
compatible with the measured product of the branching fractions 
for $B^+ \to X K^+$ and $X \to J/\psi \, \pi^+ \pi^-$ if 
${\rm Br}[X \to J/\psi \, \pi^+ \pi^-]$ is greater than about $10^{-1}$.

Our result for the ratio of the branching fractions 
for $B^0 \to X K^0$ and $B^+ \to X K^+$ is given in Eq.~(\ref{Gratio}).
It is expressed in terms of parameters $G_1$, $G_2$, $G_3$, and $G_4$
that appear in the amplitudes for $B\to \bar D^{(*)} D^{(*)} K$.
The result is independent of the binding energy $E_b$ of the $X$ 
and also independent of the crossover scale $\Lambda_\pi$.
Based on the determination of the parameters $G_i$ from 
our analysis of the Babar data, we concluded that the ratio of the 
branching fractions should be less than about $8 \times 10^{-2}$.
The suppression of $B^0 \to X K^0$ 
can be explained partly by the decays of 
$B^0$ into $D^0 \bar D^{*0} K^0$ and $D^{*0} \bar D^0 K^0$
being dominated by color-suppressed amplitudes 
and partly by heavy quark symmetry forcing these 
amplitudes to vanish at the $D D^*$ threshold.

In our analysis of the Babar data on the decays 
$B \to D^{(*)} \bar D^{(*)} K$, we made the crude 
assumption that the form factors $G_1(q^2)$, $G_2(q^2)$, $G_3(q^2)$,
and $G_4(q^2)$ are constants.  The primary reason for this assumption 
was that the available experimental information was limited to
branching fractions for the decays $B \to D^{(*)} \bar D^{(*)} K$.
Measurements of Dalitz plot distributions and invariant mass 
distributions for those decays would allow a more rigorous analysis 
that takes into account the $q^2$-dependence of the form factors.
This could be used to make 
a more precise prediction of the ratio of the branching fractions  
for $B^0 \to X K^0$ and $B^+ \to X K^+$.  Measurments of the 
invariant mass distributions for the decays $B \to D^0 \bar D^{*0} K$ 
and $B \to D^{*0} \bar D^0 K$ would be particularly valuable.
They might reveal the enhancement near the $D^0 \bar D^{*0}$
threshold that would confirm the interpretation of the $X$
as a $D D^*$ molecule.  Even without sufficient data to resolve 
the peak near the $D^0 \bar D^{*0}$ threshold, those invariant mass
distributions could be used to constrain the parameter
$\Lambda_\pi$ in our crude model of the crossover 
from the resonant distribution 
to the phase space distribution.  This could be used to sharpen 
our estimate of the branching fraction for $B^+ \to X K^+$,
since the expression in Eq.~(\ref{BrBXK}) depends quadratically 
on $\Lambda_\pi$.  Measurements of the invariant mass
distributions would also provide motivation for developing 
a more accurate model of the crossover.

The suppression of the decay $B^0 \to X K^0$ compared to
$B^+ \to X K^+$ is a nontrivial prediction of the interpretation 
of $X(3872)$ as a $D D^*$ molecule.
This prediction stands in sharp contrast to the observed pattern 
of exclusive decays of $B^0$ and $B^+$ into a charmonium $H$ plus $K$.
The ratios of the branching fractions for $B^0 \to H K^0$ and
$B^+ \to H K^+$ for the charmonium states $\eta_c$, $J/\psi$,
$\psi(2S)$, and $\chi_{c1}(1P)$ are $1.33 \pm 0.60$, 
$0.85 \pm 0.06$, $0.91 \pm 0.12$, and $0.59 \pm 0.20$, respectively.
Because charmonium is an isospin singlet
and the weak decay operators in Eqs.~(\ref{4fermion}) are also 
isospin singlets, isospin symmetry implies
that the ratio of the branching fractions  for $B^0 \to H K^0$ and
$B^+ \to H K^+$ should be equal to the ratio of the 
lifetimes $\tau[B^0]$ and $\tau[B^+]$, which is $0.921 \pm 0.014$.
The observed deviations from this lifetime ratio are all less than 
2 standard deviations.
If $X$ were an isosinglet, isospin symmetry would imply
that the ratio of the branching fractions for $B^0 \to X K^0$ and
$B^+ \to X K^+$ should also be equal to $\tau[B^0]/\tau[B^+]$.
Thus the observation of suppression 
of $B^0 \to X K^0$ relative to this prediction would 
disfavor any charmonium interpretation and support 
the interpretation of $X$ as a $D D^*$ molecule. 

S.~Nussinov, who was a coauthor of our previous paper
on the decay $B^+ \to X K^+$ \cite{Braaten:2004fk}, helped formulate 
the ideas on which this paper is based. 
We thank J.~Bendich for pointing out that a small ratio of the branching
fractions for $B^0 \to X K^0$ and $B^+ \to X K^+$ indicates a severe violation
of isospin symmetry if $X$ is a charmonium state.
This research was supported in part by the Department of Energy 
under grant DE-FG02-91-ER4069.


\end{document}